\documentclass[Physsubmission, Phys]{SciPost}

\hypersetup{
    colorlinks,
    linkcolor={red!50!black},
    citecolor={blue!50!black},
    urlcolor={blue!80!black}
}

\usepackage[bitstream-charter]{mathdesign}
\DeclareSymbolFont{usualmathcal}{OMS}{cmsy}{m}{n}
\DeclareSymbolFontAlphabet{\mathcal}{usualmathcal}

\begin{document}

\begin{center}{\Large \textbf{
Precision physics with the Proton Spectrometer and diffractive physics measurements from CMS\\
On behalf of the CMS and TOTEM collaborations\\
}}\end{center}

\begin{center}
Christophe Royon\textsuperscript{1}
\end{center}

\begin{center}
{\bf 1} The University of Kansas, Lawrence, USA\\
* christophe.royon@ku.edu
\end{center}

\begin{center}
\today
\end{center}


\definecolor{palegray}{gray}{0.95}
\begin{center}
\colorbox{palegray}{
  \begin{tabular}{rr}
  \begin{minipage}{0.1\textwidth}
    \includegraphics[width=23mm]{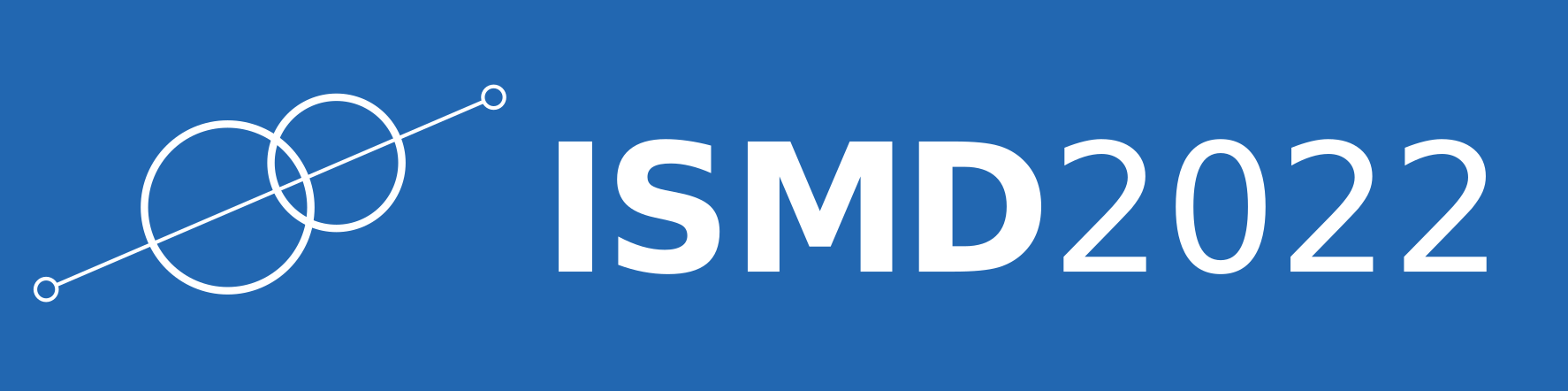}
  \end{minipage}
  &
  \begin{minipage}{0.8\textwidth}
    \begin{center}
    {\it 51st International Symposium on Multiparticle Dynamics (ISMD2022)}\\ 
    {\it Pitlochry, Scottish Highlands, 1-5 August 2022} \\
    \doi{10.21468/SciPostPhysProc.?}\\
    \end{center}
  \end{minipage}
\end{tabular}
}
\end{center}

\section*{Abstract}
{\bf
We describe recent results from CMS and TOTEM on hard diffraction, diffractive jets and jet gap jet events. We also give the first sensitivities and limits on quartic anomalous couplings and axion-like particles at high mass using the LHC as a $\gamma \gamma$ collider. The predicted sensitivities with 300 fb$^{-1}$ are better by two or three orders of magnitude compared to the more standard methods at the LHC without measuring intact protons after collision.
}

\vspace{10pt}
\noindent\rule{\textwidth}{1pt}
\tableofcontents\thispagestyle{fancy}
\noindent\rule{\textwidth}{1pt}
\vspace{10pt}

\section{Hard diffraction measurements at the LHC in the CMS collaboration with proton tagging}

At the LHC, it is possible to detect and measure single diffractive (SD) and double pomeron exchange (DPE) events by tagging one or two intact protons in TOTEM after collision, respectively. In order to get an intact proton in the final state, it is needed to have a colorless exchange that could be a photon or a pomeron (at lowest order, it corresponds to the exchange of two gluons). Photon exchanges dominate at high mass. In these measurements, we focus on low mass production and diffraction via pomeron exchange is dominating.  We define $\xi$ as the proton fractional momentum loss (the momentum fraction
of the proton carried by the pomeron) and $t$, the 4-momentum transfer squared at the proton vertex. If we assume that the pomeron has a parton structure in terms of quarks and gluons, we define $\beta$, the momentum fraction of the pomeron carried by the parton inside the pomeron that interacts, and by definition, $\beta = x / \xi$. By conservation of energy, the diffractive mass is $M=\sqrt{\xi_1 \xi_2 s}$ for DPE events where $s$ is the center-of-mass energy square.

\begin{figure}[h]
\centering
\includegraphics[width=0.4\textwidth]{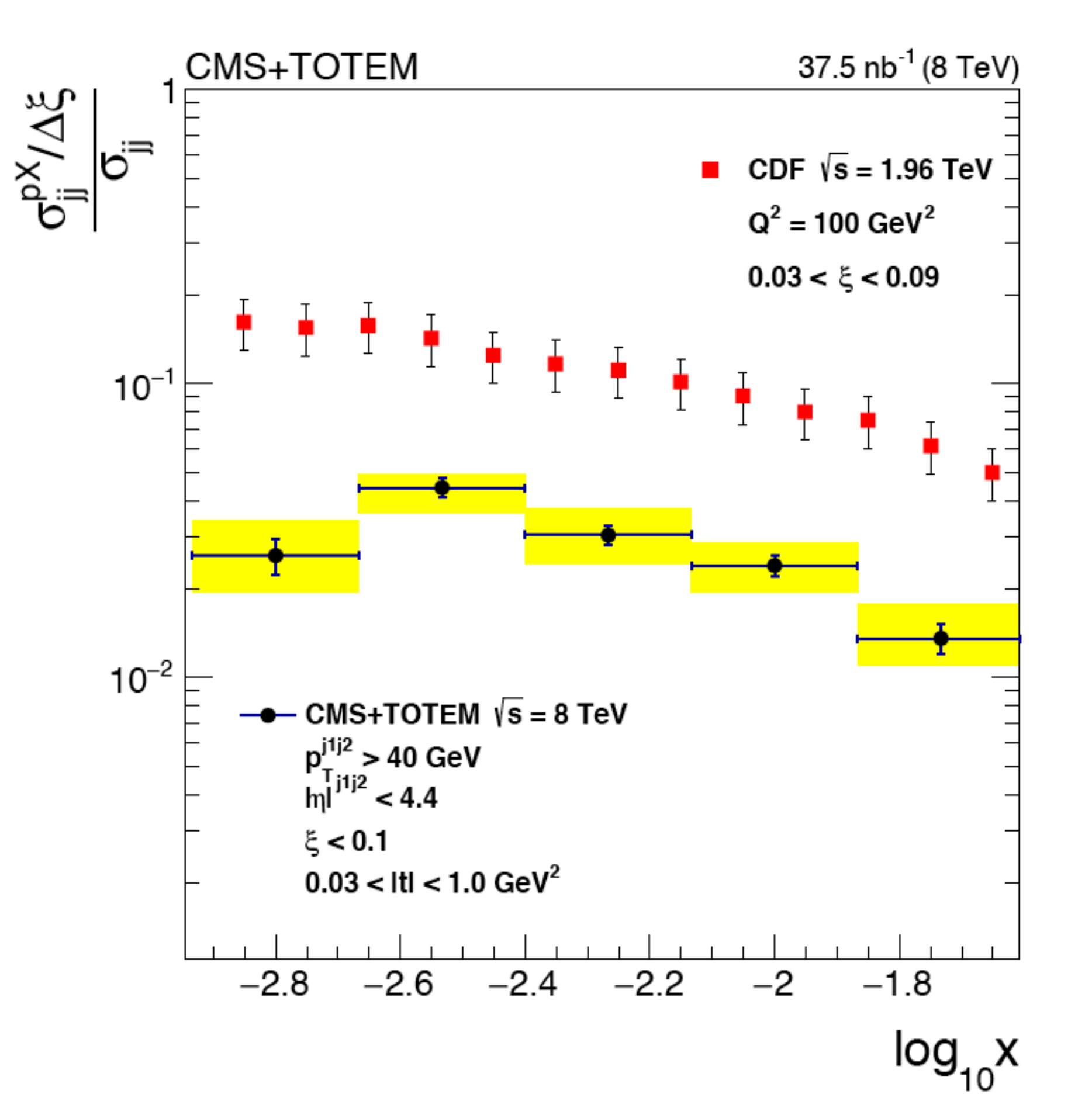}
\caption{Fraction of single diffractive jet production as a function of $\log x$ measured by the CMS and TOTEM collaboration at the 8 TeV LHC compared with an intact proton tagged to the CDF results at the Tevatron.}
\label{fig1}
\end{figure}

The advantage of the LHC machine is that we can use different values of $\beta^*$ for the beams~\footnote{$\beta^*$ describes how the beams are parallel between each other.}. High $\beta^*$ values such as 90 m or even 1.2 km lead to a good acceptance down to low $\xi$ for diffractive protons and thus low diffractive masses using vertical roman pots.
In this case, the diffractive cross section is high, and low values of accumulated luminosity are enough to measure diffractive cross sections.
On the contrary, low $\beta^*$ is used for standard high luminosity running at the LHC, that leads to a large acceptance at high diffractive mass for horizontal roman pots, and can be useful to look for beyond-standard-model anomalous coupling events. This complementarity between the different possible values of $\beta^*$ allows to cover a wide range of diffractive masses.

Many different sorts of events can be used to constrain the parton structure of the pomeron and we can quote for instance the measurements of the dijet cross section sensitive mainly to the gluon structure or $\gamma+$ jet and $W$ asymmetry to the quark density~\cite{us1,us2,us3}. As an example, the CMS and TOTEM collaborations measured the fraction of SD events as a function of $x$ at the LHC for $\sqrt{s}=8$ TeV, for a jet transverse momentum above 40 GeV, and rapidity less than 4.4, $\xi<0.1$ and $0.03 < |t|<1.0$ GeV$^2$~\cite{cmsjet} as shown in Fig.~\ref{fig1}. The data are also compared to previous measurements at the Tevatron from the CDF collaboration~\cite{cdfjet}. We notice the further suppression of SD dijets at the LHC with respect to the Tevatron due to a lower value of the survival probability that represents the probability not to have additional soft gluon exchange between the proton and the jet for instance that destroys the intact protons.

\begin{figure}[h]
\centering
\includegraphics[width=0.6\textwidth]{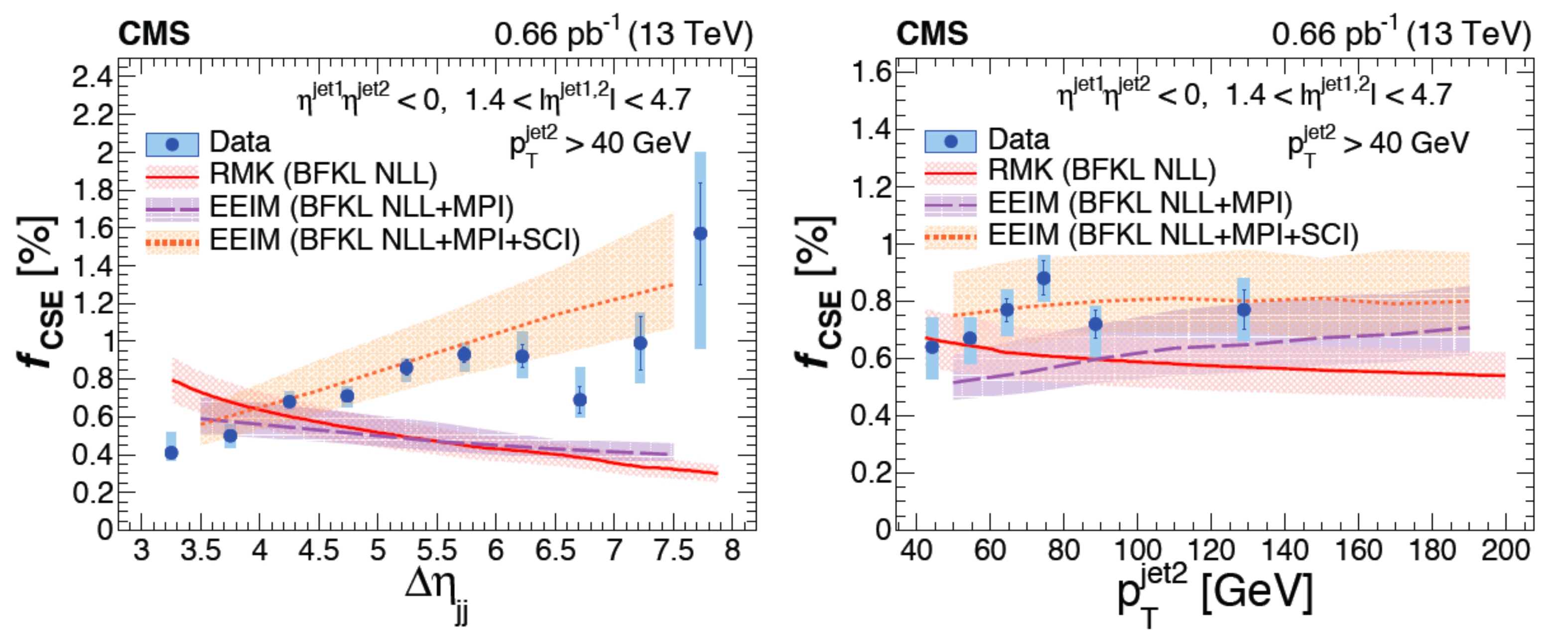}
\includegraphics[width=0.35\textwidth]{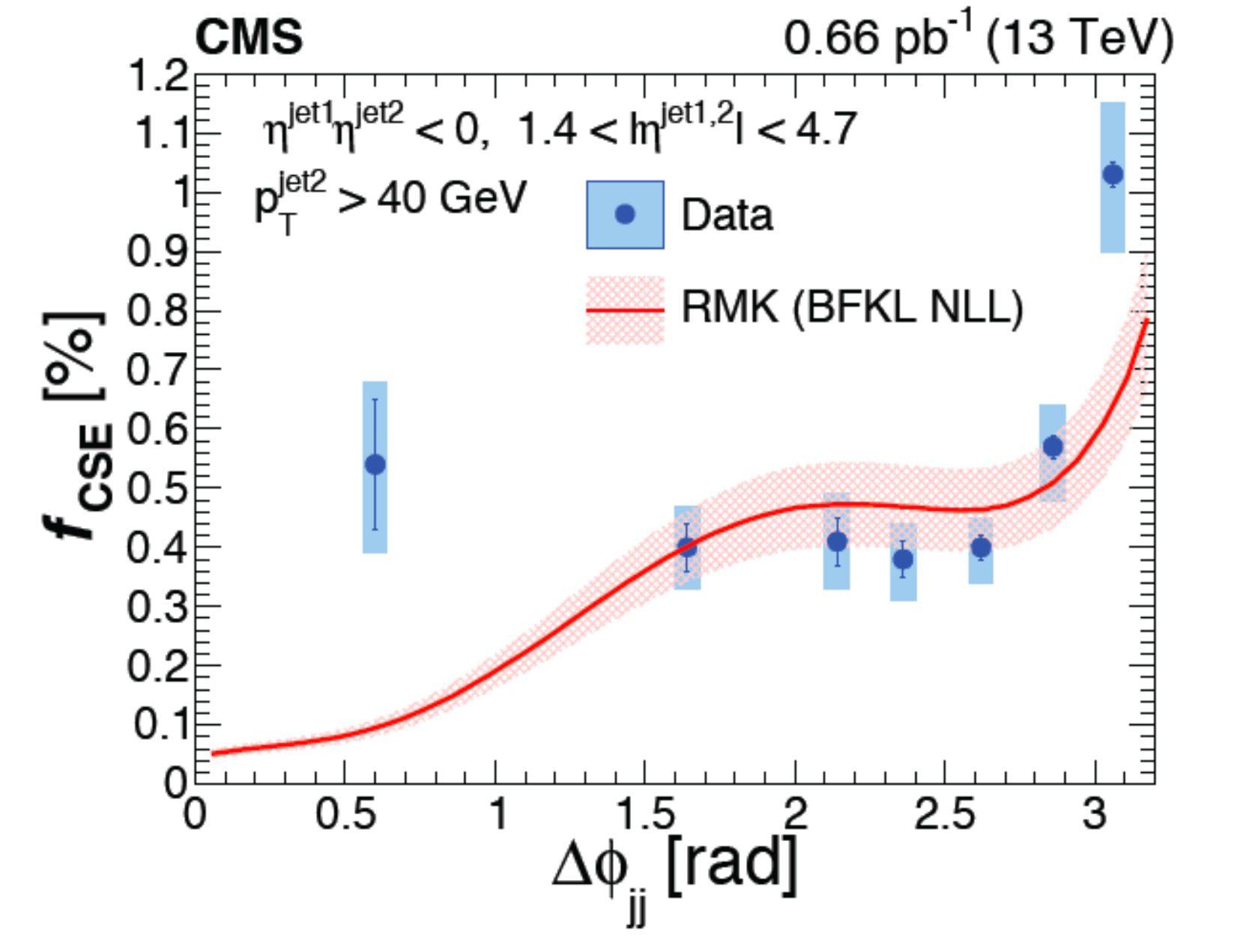}
\caption{Fraction of jet gap jet events as a function of jet $\Delta \eta$, $p_T$,
$\Delta \Phi$ compared to NLL BFKL and soft color interaction models.}
\label{fig2}
\end{figure}

\begin{figure}[h]
\centering
\includegraphics[width=0.7\textwidth]{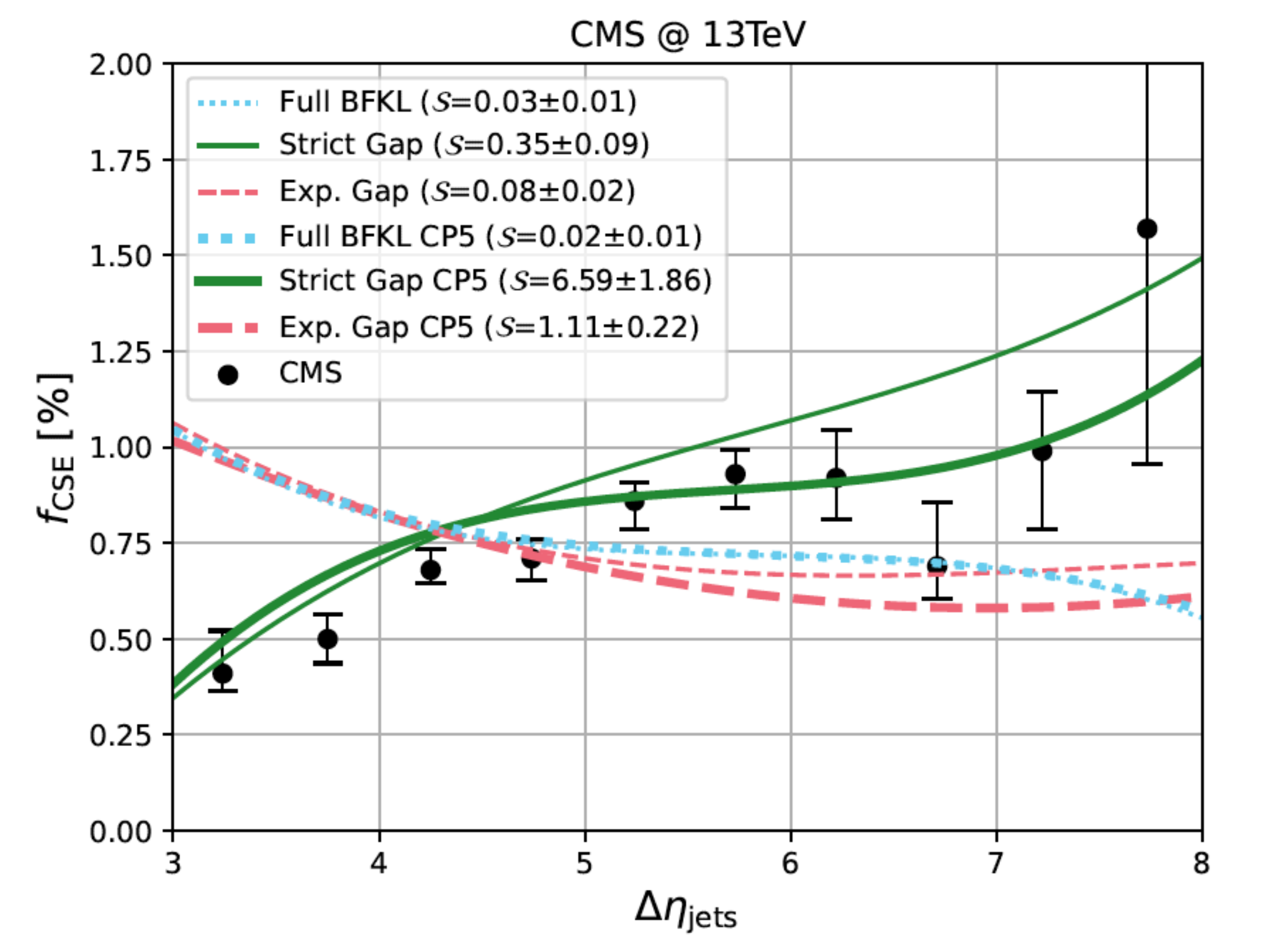}
\caption{Prediction from the ``theory" gap definition (no particle above 5 MeV in the gap $+$ ISR from pythia) and the``experimental" one (no charged particle above 200 MeV).}
\label{fig3}
\end{figure}

Another measurement that was recently performed by the CMS and TOTEM collaborations is the fraction of jet gap jet events at 13 TeV using special runs at high $\beta^*$~\cite{mueller,cmsjgj}. This kind of events is sensitive to the Balitski Fadin Kuraev Lipatov (BFKL)~\cite{bfklold1,bfklold2,bfkl} resummation effects and previous theoretical calculations have been shown to be in agreement with previous measurements at the Tevatron~\cite{jgjcyrille}. The CMS and TOTEM collaborations requested the presence of a fixed rapidity gap $-1<\eta<1$, devoid of any charged particle reconstructed using the tracking detector, and measured the fraction of jet-gap-jet events as a function of jet $\Delta \eta$, $p_T$, $\Delta \Phi$ as shown in Fig.~\ref{fig2}. Data are compared with NLL BFKL calculation, including LO impact factors~\footnote{NLL BFKL calculations including NLO impact factors are being finalized~\cite{federico}.} as implemented in PYTHIA, and with soft color interaction based models~\cite{sci}. Contrary to the Tevatron results, the agreement with the BFKL NLL calculation is reasonable for the $p_T$ and $\Delta \phi$ data but poor for the $\Delta \eta$ measurement that show an increasing distribution in data and slowly decreasing one in the calculation. It was recently noticed that the NLL predictions are very sensitive to the gap definition~\cite{jgjmuenster}.  We show in Fig.~\ref{fig3} the difference in BFKL NLL calculation between the ``theory" gap definition (no particle above 5 MeV in the gap) and the ``experimental" one (no charged particle above 200 MeV) as implemented in PYTHIA~\cite{pythia}. It seems that PYTHIA predicts too much radiation especially in initial state radiation that has tendency to fill in the gap between the jets. The amount of initial state radiations should be further tuned using additional samples such as $J/\psi$-gap-$J/\psi$ events  or the production of other  quarkonia  pairs.

Using the same sample of events, the CMS and TOTEM collaborations observed for the first time jet gap jet events in diffraction, requesting one intact proton to be measured in the TOTEM roman pots~\cite{cmsjgj} as suggested previously~\cite{jgjdiff}. It leads to very clean events for jet gap jets since MPI are suppressed and might be the ``ideal" way to probe BFKL resummation effects. 11 events were observed with $\sim 0.7$ pb$^{-1}$ and the fraction of jet gap jet events with a tagged proton is $0.66 \pm 0.01$(stat.)$^{0.06}_{0.09}$(syst.)\% larger than for the inclusive sample. The double ratio of the fraction of jet-gap-jet events in diffraction divided by the inclusive one is $0.34 \pm 0.08$(stat.)$^{0.12}_{-0.13}$(syst.). It suggests that a gap is more likely to form or survive in the presence of another gap or an intact proton.
This measurement would benefit from more statistics, about 10 pb$^{-1}$ for SD and 100 pb$^{-1}$ for DPE to perform a more differential measurement.

\

\section{The LHC as a $\gamma \gamma$ collider}

In this section, we will discuss the exclusive production of $\gamma \gamma$, $WW$, $ZZ$ and $t \bar{t}$ at high luminosity in standard runs at the LHC, and we assume that we can detect and measure the intact protons after interaction in the Proton Precision Spectrometer (PPS). As we mentioned in the previous section, the acceptance starts at high $\xi$ for low $\beta^*$ running for proton tagging, and the diffractive mass acceptance starts at about 450 GeV. It means that $\gamma \gamma$, $WW$, $ZZ$ and $t \bar{t}$ exclusive productions are dominated by photon-exchange events and the QCD processes can be neglected~\cite{gammagamma5,gammagamma6}. The LHC can thus be considered as a $\gamma \gamma$ collider. It is also worth noticing that the exclusive production of diphotons at low mass can be studied in 
$Pb Pb$ interactions since the exclusive $\gamma \gamma$ cross section is enhanced by a factor  $Z^4$~\cite{axion1}.

\subsection{Semi-exclusive production of dileptons}

We first consider the QED semi-exclusive production of dileptons, $ee$, and $\mu \mu$.  We request at least one proton to be measured in PPS~\footnote{Requesting both protons to be intact and detected in PPS would lead to few events to be observed because of the low value of the cross section of exclusive dilepton production.}. 17 (resp. 23) events are found with protons in the PPS acceptance and 12 (resp. 8) show a
less than 2$\sigma$ matching in the $\mu \mu$ (resp. $ee$) channel~\cite{cmsexcl}.
The significance for observing 20 events for a
background of $3.85$ ($1.49\pm 0.07 (stat) \pm 0.53 (syst)$ for $\mu \mu$ and 
$2.36\pm 0.09 (stat) \pm 0.47 (syst)$ for $ee$) is found to be larger than 5$\sigma$. No signal candidate event is found at high mass with both protons tagged  (the two dielectron events in the acceptance region are compatible with a pile up contamination since 2.36 background events are expected).

\subsection{$\gamma \gamma$ exclusive production}

We will first discuss the general method to look for exclusive $\gamma \gamma$, $WW$, $ZZ$, $\gamma Z$ $t \bar{t}$ production with tagged protons at the LHC considering as an example the exclusive $\gamma \gamma$ production.
Let us first notice that it is possible to obtain a larger number of events  than expected in SM due to extra-dimensions, composite Higgs models, axion-like particles production.  Quartic photon anomalous couplings can appear via loops of new particles coupling to photons or via resonances decaying into two photons~\cite{gammagamma,gammagamma2,gammagamma3,gammagamma4}. 

\begin{figure}[h]
\centering
\includegraphics[width=0.8\textwidth]{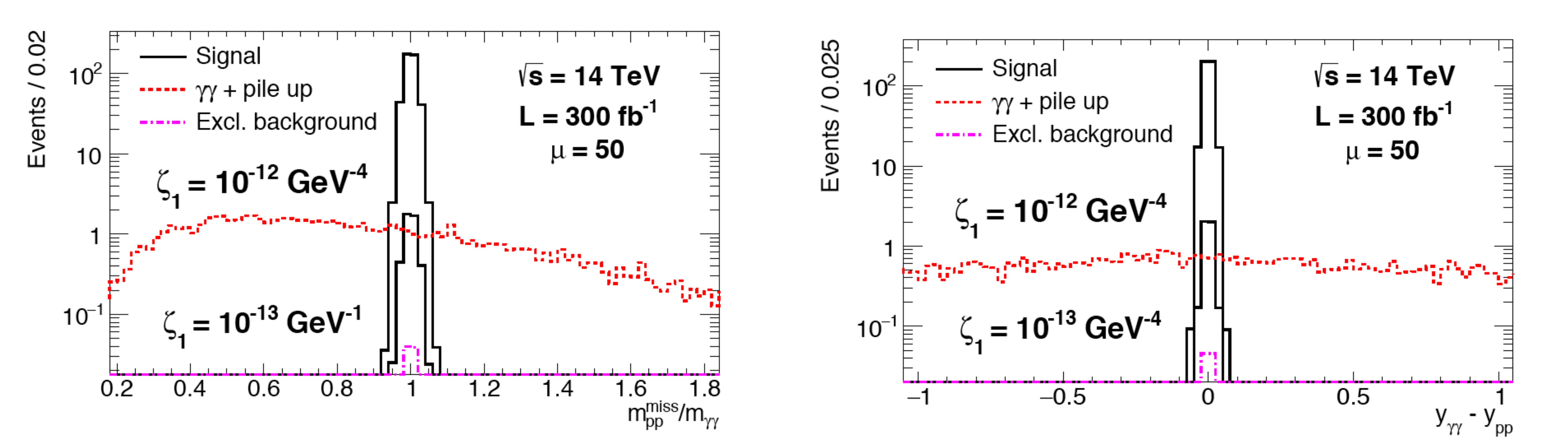}
\caption{Ratio between the diproton and diphoton mass and difference between  the diproton and diphoton rapidity for signal and background.}
\label{fig3b}
\end{figure}

The general idea of looking for exclusive $\gamma \gamma$ production is first to request as a preselection high transverse momentum photons, back-to-back, and a high diphoton mass (we recall that the diffractive mass acceptance starts at about 450 GeV for double proton tagged events).
The advantage of exclusive events is that we detect and measure all particles produced in the final state after interaction, namely the two photons and the two protons. We are in a very clean situation like at LEP. Energy and momentum conservation allow to get a very efficient background suppression. Let us first note that the only remaining background after preselection is the inclusive production of two photons and additional protons originating from pile up. The suppression of the pile up background is illustrated in Fig.~\ref{fig3b} where we display the ratio between the diproton and diphoton mass and difference between  the diproton and diphoton rapidities~\footnote{The diproton mass and rapidity are computed using $M=\sqrt{\xi_1 \xi_2 s}$ and $y=0.5 \log \xi_1/\xi_2$ where $\xi_1$, and $\xi_2$ are the proton momentum losses.} for signal and pile up background. Since, for pile up events, the two intact protons are not related to the diphoton production, this is a very efficient method to remove the background~\cite{gammagamma5,gammagamma6}. In the case where not all produced particles can be measured in CMS (for instance the case of exclusive $WW$ production where the $W$s decay leptonically and neutrinos are produced), the use of fast timing detectors is crucial to constrain the intact protons to originate from the same vertex as the $WW$ (or $t \bar{t}$) production~\cite{ttbar}.

\begin{figure}[h]
\centering
\includegraphics[width=0.5\textwidth]{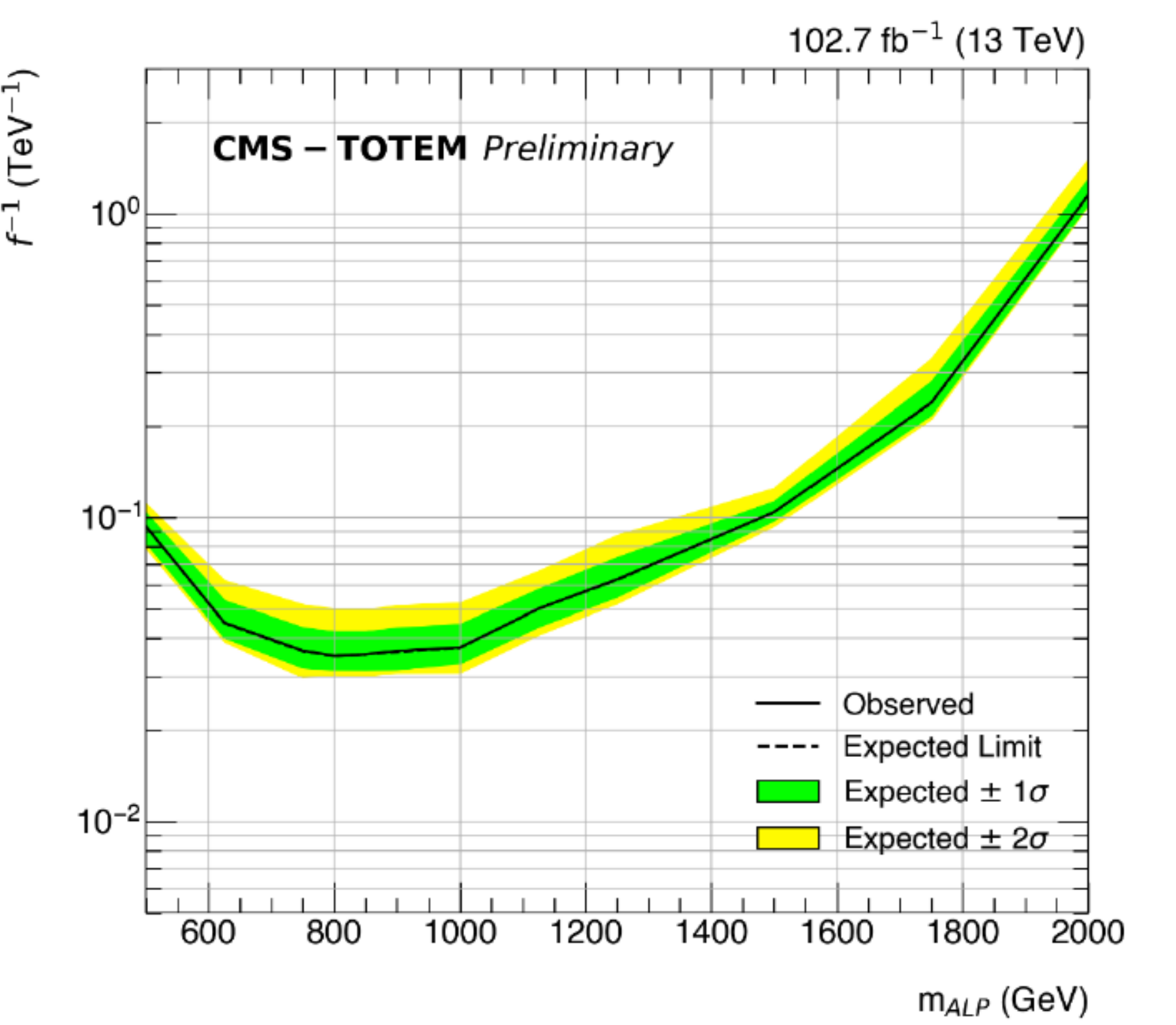}
\caption{First limits on the production of axion-like particles in the coupling versus mass plane using the exclusive production of two photons at the LHC.}
\label{fig4}
\end{figure}

The CMS and TOTEM collaborations looked for the exclusive diphoton production at high mass with intact protons at high luminosity for the first time requesting back-to-back photons, a high diphoton mass ($m_{\gamma \gamma}>350$ GeV), matching in rapidity and mass with the  diproton information. No signal was found and It leads to the
first limits on quartic photon anomalous couplings, $|\zeta_1|<2.9~10^{-13}$ GeV$^{-4}$, $|\zeta_2|<6.~10^{-13}$ GeV$^{-4}$ at 95\% CL with about a luminosity of 10 fb$^{-1}$~\cite{gammagamma8}. The results were recently updated with a luminosity of
102.7 fb$^{-1}$, that leads to limits of $|\zeta_1|<7.3~10^{-14}$ GeV$^{-4}$, $|\zeta_2|<1.5~10^{-13}$ GeV$^{-4}$ at 95\% CL~\cite{gammagamma7}. It also leads to the first limits on axion-like particles decaying into two photons as shown in Fig.~\ref{fig4} in the coupling versus axion mass plane.

\subsection{Exclusive production of $W$ and $Z$ boson pairs}

The $WW$ and $ZZ$ exclusive productions represent a clean test of the electroweak part of the SM. One can look for the SM production of $WW$ and $ZZ$ using the leptonic decays of the $W$ bosons because the hadronic channel suffers from the high dijet background~\cite{ww}. On the contrary, the search for anomalous $\gamma \gamma WW$ or $\gamma \gamma ZZ$ benefits from the pure hadronic decays of the $W$ and $Z$ bosons since the $W$ and $Z$ are produced at high $WW$ and $ZZ$ masses where the dijet background is smaller~\cite{ww}.

\begin{figure}[h]
\centering
\includegraphics[width=0.4\textwidth]{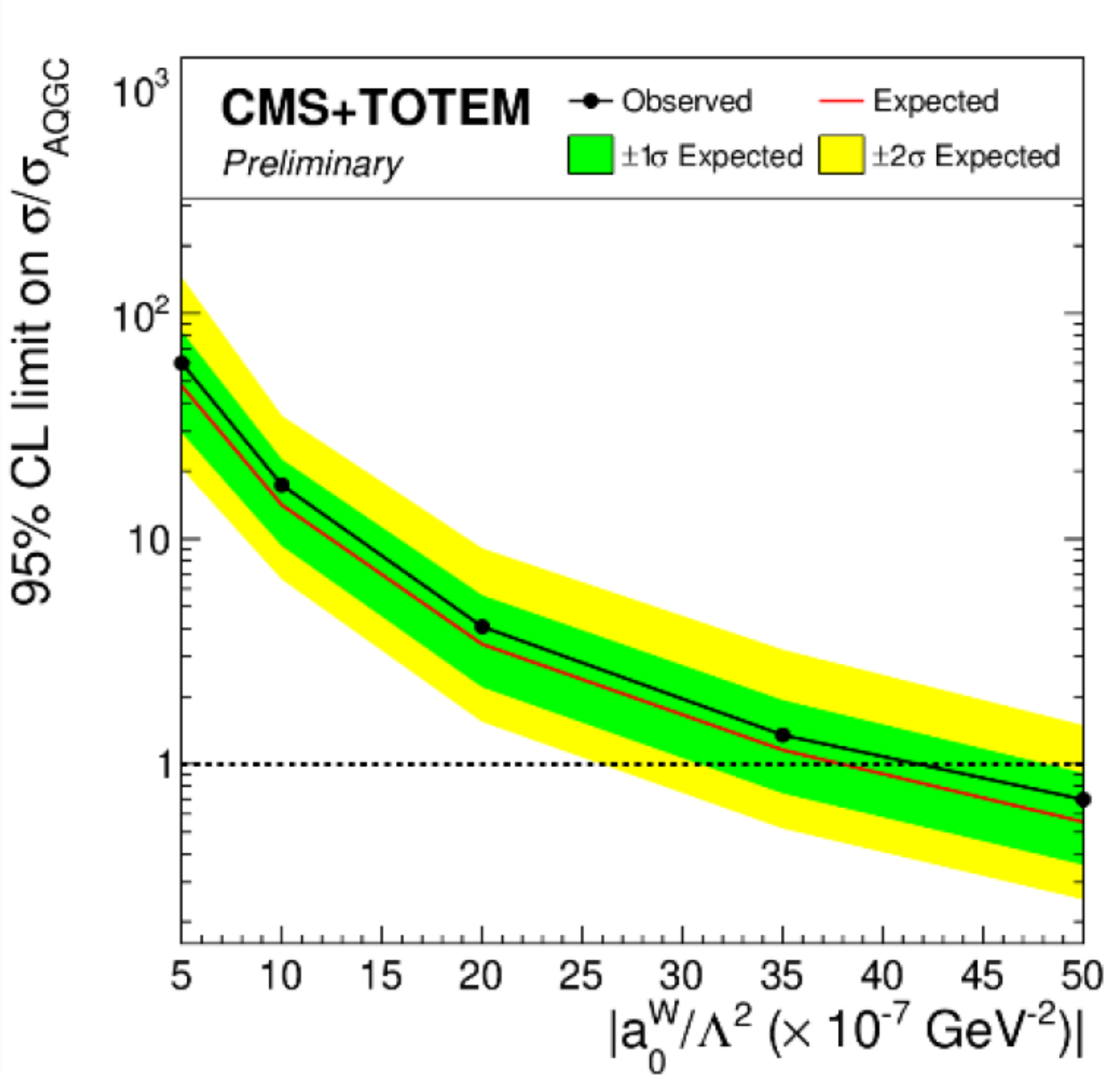}
\caption{Limits on quartic anomalous $\gamma \gamma WW$ $a_0^W$ coupling.}
\label{fig5}
\end{figure}

The CMS and TOTEM collaborations  looked for anomalous exclusive $WW$ production in the full hadronic decays of the $W$ bosons, requesting the reconstruction of two ``fat" jets of radius 0.8, with a jet $p_T>200$ GeV, and 1126$<m_{jj}<$2500 GeV, the jets being back-to-back ($|1-\phi_{jj}/\pi|<0.01$). 
As usual, the signal region is defined by the correlation between the central $WW$ system and the proton information. No signal was found and the limits on the SM cross section  are $\sigma_{WW}<67$fb, $\sigma_{ZZ}<43$fb for $0.04<\xi<0.2$ at 85\% CL~\cite{ww1}. 
In addition, new limits were derived on quartic anomalous couplings, $a_0^W/\Lambda^2 < 4.3~10^{-6}$ GeV$^{-2}$, $a_C^W/\Lambda^2 < 1.6~10^{-5}$ GeV$^{-2}$, $a_0^Z/\Lambda^2 < 0.9~10^{-5}$ GeV$^{-2}$, $a_C^Z/\Lambda^2 < 4.~10^{-5}$ GeV$^{-2}$ with 52.9 fb$^{-1}$ at 95\% CL. As an example, the limit on $a_0^W/\Lambda^2 $ is shown in Fig.~\ref{fig5}.

It is worth noticing that an additional search for the exclusive production of $\gamma Z$ has now started. This is specially interesting that it is possible to look both for hadronic and leptonic decays of the $Z$ boson~\cite{gammaz}.

\subsection{Exclusive production of $t \bar{t}$}
The CMS and TOTEM collaborations looked for exclusive $t \bar{t}$ production in the dilepton and semileptonic channels~\cite{ttbar1}. In the dilepton channels, two leptons with $p_T>$30, 20 GeV, with a dilepton mass outside the $Z$ boson mass were requested, with two $b$-jets and one proton tagged on each side of CMS, and for the semileptonic channel, one lepton, two $b$-jets, two non $b$-jets, and one proton on each side of CMS.
In order to reduce the relatively high background (due to the fact that the mass and rapidity matchings are not that efficient due to the presence of neutrinos) constraints based on the $W$ and top masses were used. No signal was found and a limit was put on the exclusive production of $t \bar{t}$ $\sigma^{excl.}_{t \bar{t}} < 0.6$ pb. 

The search for exclusive $t \bar{t}$ will strongly benefit from the use of fast timing detectors to reduce the pile up background (requesting that the protons originate from the same vertex as the $t \bar{t}$)~\cite{ttbar}.

\subsection{Exclusive production of $Z+X$ and $\gamma + X$}
\begin{figure}[h]
\centering
\includegraphics[width=0.8\textwidth]{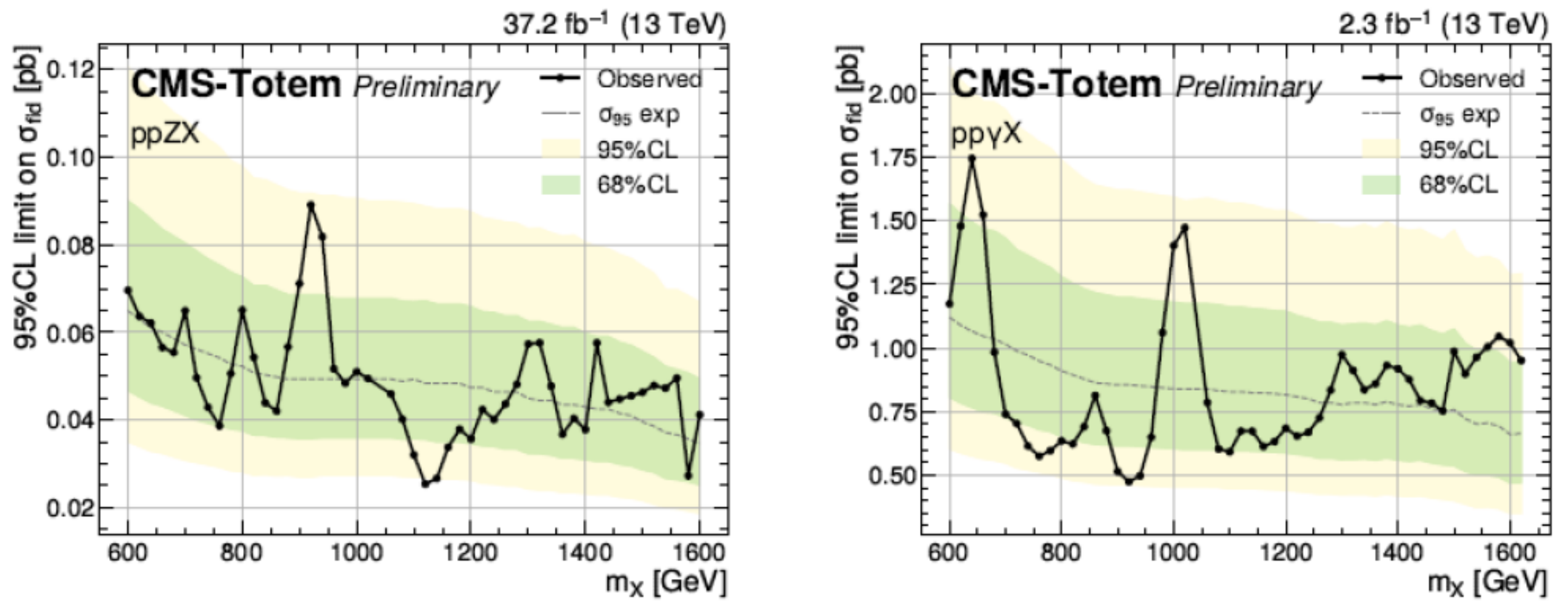}
\caption{95\% CL limit for $Z+X$ and $\gamma +X$ production as a function of $m_X$.}
\label{fig6}
\end{figure}

Using the matching in mass and rapidity between the particles measured in the main CMS detector and the proton information in PPS, the CMS and TOTEM collaborations looked for the exclusive production of $\gamma+X$ or $Z+X$, the associated production of a photon or a $Z$ boson plus any particle $X$ that might disappear in the detector (any invisible particle such as the lightest SUSY particle as an example). No signal was found and the 95\% CL limit is given in Fig.~\ref{fig6} for $Z+X$ and $\gamma +X$ events (please note that the luminosity is not the same for both searches because of the different triggers that were used)~\cite{zx}. 
This model-independant analysis will clearly benefit from higher luminosity at the LHC.

\section{Conclusion}
In this report, we first discussed the CMS and TOTEM results on hard diffraction such as the measurement of diffractive jets at 8 TeV. At 13 TeV, the fraction of jet gap jet events was measured and such events were observed for the first time in diffraction with one proton tagged in the final state. The search for exclusive production of $\gamma \gamma$, $WW$, $ZZ$, $t \bar{t}$, $\gamma+X$ and $Z+X$ via photon exchanges was performed by the CMS and TOTEM collaborations requesting the presence of two intact protons on each side of the CMS detector. This leads to be the best sensitivities to anomalous quartic couplings or to the production of axion-like particles at high masses.

\nolinenumbers

\end{document}